\documentclass[aps,prc,preprint,groupedaddress,amsmath]{revtex4-1}

\usepackage{graphicx}
\usepackage{braket}
\usepackage{bm}
\usepackage{longtable}

\usepackage{amsmath}	
\begin{document}

\title{Inversion doublets of reflection-asymmetric clustering in $^{28}$Si\\
   and their isoscalar monopole and dipole transitions}

\author{Y. Chiba}
\affiliation{Department of Physics, Hokkaido University, 060-0810 Sapporo, Japan}
\author{Y. Taniguchi}
\affiliation{Faculty of Health Sciences, Nihon Institute of Medical Science,
Moroyama, Saitama 350-0435, Japan} 
\author{M. Kimura}
\affiliation{Department of Physics, Hokkaido University, 060-0810 Sapporo, Japan}
\affiliation{Nuclear Reaction Data Centre, Faculty of Science, Hokkaido University, Sapporo
060-0810, Japan} 

\date{\today}

\begin{abstract}
\begin{description}
\item[Background] 
 Various cluster states of astrophysical interest are expected to exist
 in the excited states of $^{28}{\rm Si}$. However, they have not been identified firmly,
 because of the experimental and theoretical difficulties.
\item[Purpose]
 To establish the $^{24}$Mg+$\alpha$, $^{16}$O+$^{12}$C and $^{20}$Ne+2$\alpha$ cluster
 bands, we theoretically search for the negative-parity cluster bands that are paired with the
 positive-parity bands to constitute the inversion doublets.
 We also offer the isoscalar monopole and dipole transitions as a promising probe for the
 clustering. We numerically show that these transition strengths from the ground state to the
 cluster states are very enhanced.
\item[Method]
 The antisymmetrized molecular dynamics with Gogny D1S effective interaction is employed to
 calculate the excited states of $^{28}{\rm Si}$. The isoscalar monopole and dipole transition
 strengths are directly evaluated from wave functions of the ground and excited states.
\item[Results]
  Negative-parity bands having $^{24}$Mg+$\alpha$ and $^{16}$O+$^{12}$C cluster configurations
  are obtained in addition to the newly calculated $^{20}$Ne+2$\alpha$ cluster bands.
  All of them are paired with the corresponding positive-parity bands to constitute the
  inversion doublets with various cluster configurations. The calculation show that the
  band-head of the $^{24}$Mg+$\alpha$ and $^{20}$Ne+2$\alpha$ cluster bands are strongly excited
  by the isoscalar monopole and dipole transitions.
\item[Conclusions]
  The present calculation suggests the existence of the inversion doublets with the 
  $^{24}$Mg+$\alpha$, $^{16}$O+$^{12}$C and $^{20}$Ne+2$\alpha$ configurations.
  Because of the enhanced transition strengths, we offer the isoscalar monopole and dipole
  transitions as good probe for the $^{24}$Mg+$\alpha$ and $^{20}$Ne+2$\alpha$ cluster bands.
\end{description}
\end{abstract}

\pacs{Valid PACS appear here}
\maketitle

\section{Introduction} \label{sec:1}
Compared to light $p$-shell nuclei where the $\alpha$ clustering is prominent in the ground and
excited states \cite{Fujiwara1980}, the structure of the $sd$-shell nuclei is more complicated
but much richer. The formation of the mean-field in the low-lying states, and its coexistence and
mixing with the $\alpha$, Carbon and Oxygen clustering yield various kinds of nuclear structures
\cite{Kimura2004,Kimura2004a,Taniguchi2009}. $^{28}{\rm Si}$ is a typical example in the mid $sd$-shell nuclei. Its ground-state band has
oblate deformed mean-field structure, while the rotational band built on the $0^+_3$ state is
prolately deformed suggesting the shape coexistence 
\cite{DasGupta1967,Glatz1981,Glatz1981a,Sheline1982}. It is believed 
that the $\beta$ vibration of the ground-state  band yields another rotational band built on the
$0^+_2$ 
state \cite{DasGupta1967,Sheline1982}. Furthermore, the possible existence of the superdeformed
(SD) state is predicted by theoretical studies 
\cite{Cseh1992,Ohkubo2004,Maruhn2006,Taniguchi2009,Ichikawa2011,Darai2012}
and the experimental candidates were observed \cite{Kolata1988,Beck1985,Eswaran1993,Jenkins2012}. 

In addition to these mean-field dynamics, $^{28}{\rm Si}$ offers a rich variety of clustering. 
Many candidates of $^{24}{\rm Mg}+\alpha$ cluster states are observed at more than 10 MeV above
the ground state \cite{Maas1978,Cseh1982,Tanabe1983,Kubono1986,Artemov1990}. 
The existence of the $^{12}{\rm C}+{}^{16}{\rm O}$ molecular 
resonance above $E_x\simeq 30$ MeV is also well known 
\cite{Stokstad1972,Frohlich1976,Charles1976,James1976,Shawcross2001,Ashwood2001,Goasduff2014}.
In addition to these highly excited cluster states, the possible clustering of the above-mentioned
low-lying mean-field states has attracted many interests and has long been discussed  
\cite{Ichikawa2011,Taniguchi2009,Baye1976,Baye1977,Kato1985,Kondo2004}. For example, by
the antisymmetrized molecular dynamics (AMD) study \cite{Taniguchi2009}, it was pointed out that 
the ground-state band and SD
band have large overlap with the $^{24}{\rm Mg}+\alpha$ cluster configurations, while the prolate
band largely overlaps with the $^{12}{\rm C}+{}^{16}{\rm O}$ configuration. A similar discussion
was also made by the algebraic cluster model \cite{Darai2012,Cseh1983,Cseh1992,Hess2004}. 
It is also noted that the $\alpha$ and
Carbon cluster states at their decay thresholds are of astrophysical interest and importance,
because they are closely related to and seriously affect to the stellar processes such as the He-
and C-burning. Thus, the clustering phenomena in $^{28}{\rm Si}$ is rich in variety and
has scientific importance.

Despite of these strong interest and the continuous experimental and theoretical 
efforts, the clustering systematics in $^{28}{\rm Si}$ is still ambiguous.
Theoretically, the description of various kinds of cluster configurations within a single
theoretical framework is not easy. In particular, the coexistence and mixing of mean-field and
cluster configurations makes the theoretical description hard, although it enriches the
clustering phenomena in $^{28}{\rm Si}$. Experimentally, the direct production of the cluster
states by the transfer and radiative capture reactions are difficult, because of their small
cross sections and very high level density. These difficulties prevent us from the
understanding and establishment of the clustering systematics in $^{28}{\rm Si}$.

To overcome these difficulties, we present the result of the AMD calculation to illustrate the
clustering systematics and suggest the isoscalar (IS) monopole and dipole transitions as
promising probe for clustering. We focus on the following two points.
The first is the negative-parity cluster bands and the identification of the inversion doublets.
The previous AMD study \cite{Taniguchi2009} investigated the structure of the positive-parity
bands, and found that three positive-parity bands, the ground-state, prolate deformed and
SD bands, have large overlap with the triaxial deformed $^{24}{\rm Mg}+\alpha$,
the $^{16}{\rm O}+{}^{12}{\rm C}$ and the axially symmetric $^{24}{\rm Mg}+\alpha$ cluster
configurations, respectively. If these bands really have clustering aspects, they must be
accompanied by the negative-parity bands to constitute the inversion doublets, because of their
reflection-asymmetric intrinsic configurations \cite{Horiuchi1968}. Therefore, the identification
of the inversion doublet is rather essential to establish the clustering systematics in 
$^{28}{\rm Si}$, and in this study, we extend our survey to the negative-party states in this
study. 

The second is the IS monopole and dipole transitions which are expected to strongly populate the
cluster inversion doublets. In this decade, the IS monopole transition attracts much interest as
a good probe for the $0^+$ cluster states
\cite{Suzuki1987,Kawabata2007,Kanada-Enyo2007,Yamada2008,Horiuchi2012,Chiba2015}. In addition to
this, recently, it was also 
suggested that the IS dipole transition is a good prove for the $1^-$ cluster states
\cite{Chiba2016}. Therefore, the combination of the IS monopole and dipole transitions is expected
to be a powerful tool to identify the inversion doublets mentioned above. Indeed, in
Refs. \cite{Yamada2008,Chiba2016},  by assuming that the ground states is a pure $SU(3)$ shell
model state   \cite{Elliott1958,Elliott1958a} and the excited states are ideal cluster states, it
was analytically proved that the IS monopole and dipole transition from the ground state to the 
excited cluster states are as strong as the single-particle estimates. However, in reality, the
ground state of $^{28}{\rm Si}$ deviates from a pure $SU(3)$ shell model state because of the
strong influence of the spin-orbit interaction. Furthermore, in the excited states, the cluster
configurations are mixed with the mean-field configurations. Therefore, the numerical calculations
by the reliable nuclear models are indispensable for the quantitative discussions.

In this paper, we show that the various kinds of the cluster bands including the newly found
$^{20}{\rm Ne}+{}^{8}{\rm Be}$ band appear in the negative-parity and paired with the
positive-parity bands to constitute the inversion doublets. The analysis of the wave function
shows that the ground state has the duality of oblate deformed mean-field and the clustering of
$^{24}{\rm Mg}+\alpha$ and $^{20}{\rm Ne}+{}^{8}{\rm Be}$ despite of the strong influence of
the spin-orbit interaction. Because of this duality, by the IS monopole and dipole transitions,
the inversion doublets having the $^{24}{\rm Mg}+\alpha$ and $^{20}{\rm Ne}+{}^{8}{\rm Be}$
configurations are excited as strong as the single-particle estimates. Hence, it is concluded
that the IS monopole and dipole transitions are regarded as promising probe for these
clustering.

This article is organized as follows. In the next section, the framework of AMD is briefly
explained. The cluster $S$-factor is also introduced as a measure of the clustering in the
ground and excited states. In the Sec. \ref{sec:3}, the intrinsic wave function obtained
by the energy variation and the energy spectrum are presented. The systematics of the clustering
in $^{28}{\rm Si}$ is summarized. The relationship between the clustering in the excited states
and the IS monopole and dipole transitions is discussed in the Sec. \ref{sec:4}. The final
section summarizes this work.

\section{Framework}  \label{sec:2}
Here, the framework of AMD is briefly explaind, and readers are directed to Refs. 
\cite{Kanada-Enyo2003,Kanada-Enyo2010,Kanada-Enyo2012} 
for the detailed explanation.
\subsection{Hamiltonian and variational wave function}
The $A$-body microscopic Hamiltonian used in this study reads,
\begin{align}
  {H} = \sum_{i}^{A}{{t}_i} - {t}_{c.m.} + \sum_{i<j}^{A}{{v}_{ij}^{NN}} +
 \sum_{i<j}^{A}{{v}_{ij}^{Coul}}. 
  \label{hamiltonian}
\end{align}
Here, ${t}_i$ is kinetic energy of $i$-th nucleon. ${t}_{c.m.}$ is the center-of-mass kinetic
energy which is exactly subtracted without approximation in the AMD framework.  We employ Gogny
D1S interaction \cite{Berger1991} as an effective nuclear interaction $v^{NN}$. The Coulomb
interaction $v^{Coul}$ is approximated by a sum of seven Gaussians. 

The intrinsic wave function of AMD is an antisymmetrized product of nucleon wave packets
$\varphi_i$, 
\begin{align}
 \Phi_{int} &= {\mathcal A}\left\{\varphi_1\varphi_2 \cdots \varphi_A \right\},
\end{align}
where the nucleon wave packet is a direct product of the deformed Gaussian spatial part,
spin ($\chi_i$) and isospin ($\xi_i$) parts,   
\begin{align}
 \varphi_i({\bm r}) &= \phi_i({\bm r}) \chi_i \xi_i, \label{eq:singlewf}\\
 \phi_i({\bm r}) &= \exp\biggl\{-\sum_{\sigma=x,y,z}\nu_\sigma
 \Bigl(r_\sigma -\frac{Z_{i\sigma}}{\sqrt{\nu_\sigma}}\Bigr)^2\biggr\}, \\
 \chi_i &= a_i\chi_\uparrow + b_i\chi_\downarrow,\quad
 \xi_i = {\rm proton} \quad {\rm or} \quad {\rm neutron}.\nonumber
\end{align}
The centroids of the Gaussian wave packet $\bm Z_i$, the direction of nucleon spin $a_i, b_i$,
and the width parameter of the deformed Gaussian $\nu_\sigma$ are the variational parameters
\cite{Kimura2004a}. 

Before the energy variation, the intrinsic wave function is projected to the eigenstates of the
parity, 
\begin{align}
 \Phi^\pi=\frac{1+\pi{P}_x}{2}\Phi_{int}, \quad \pi=\pm.
\end{align}
Using this wave function, the variational energy is defined as,
\begin{align}
 E^\pi = \frac{\braket{\Phi^\pi|H|\Phi^\pi}}{\braket{\Phi^\pi|\Phi^\pi}} + V_c.
\end{align}
By the frictional cooling method, above-mentioned variational parameters are
determined so that $E^\pi$ is minimized. Here $V_c$ is the potential which imposes the
constraint on the variational wave function. In this study, we introduce two different
constraint potentials. The first is the $\beta\gamma$-constraint which is imposed
on the quadrupole deformation of the variational wave function, 
\begin{align}
 V_c =  v_\beta(\braket{\beta} - \beta_0)^2  + v_\gamma (\braket{\gamma} - \gamma_0)^2,
\end{align}
where $\braket{\beta}$ and $\braket{\gamma}$ are the quadrupole deformation parameters of the
intrinsic wave function defined in Ref. \cite{Kimura2012}, and $v_\beta$ and $v_\beta$ are
chosen large enough that $\braket{\beta}$ and $\braket{\gamma}$ are close to $\beta_0$ and
$\gamma_0$ after the  frictional cooling.

Another constraint is $d$-constraint \cite{Taniguchi2004} which is  imposed on the distance
between  
{\it quasi clusters},
\begin{align}
 V_c = v_d (\braket{d^2} - d_0^2)^2. 
\end{align}
Similar to the $\beta\gamma$-constraint, $v_d$ is chosen so that the squared distance between
quasi clusters $\braket{d^2}$ is close to $d_0^2$ after the frictional cooling. 
The squared distance between quasi clusters is defined as follows. First, we select nucleons which
belongs to each quasi cluster. For example, in the case of the $\alpha+^{24}{\rm Mg}$
configuration, we choose 4 nucleon wave packets which belong to $\alpha$ cluster and regard
remaining 24 wave packets as belonging to $^{24}{\rm Mg}$ cluster. Then, we define the
center-of-mass of these quasi-clusters as  
\begin{align}
 \bm R_{\alpha} = \frac{1}{4}\sum_{i\in \alpha}\Re(\bm Z_i),\quad
 \bm R_{^{24}{\rm Mg}} = \frac{1}{24}\sum_{i\in ^{24}{\rm Mg}}\Re(\bm Z_i),
\end{align}
and define $\braket{d^2}$ as the squared distance between those centers-of-mass,
\begin{align}
 \braket{d^2} = |\bm R_\alpha - \bm R_{^{24}{\rm Mg}}|^2.
\end{align}
Then we find the wave function which yields the minimum energy for given value of $\braket{d^2}$
by the energy minimization. By applying this constraint, various kinds of cluster configurations
have been studied \cite{Taniguchi2004,Taniguchi2007,Taniguchi2009,Taniguchi2014}. In the present
study, we calculated $\alpha+{}^{24}{\rm Mg}$,  $^{12}{\rm C}+{}^{16}{\rm O}$ and 
$^{8}{\rm Be}+{}^{16}{\rm O}$ cluster configurations. It is 
noted that the $d$-constraint imposes the constraint on the distance between the quasi clusters,
but do not on their internal structure. As a result, when the inter-cluster distance is small, the
clusters are strongly polarized to gain more binding energy. On the other hand, when the distance
is sufficiently large, the clusters are in their ground states. In other words, this constraint
smoothly connects the mean-field and cluster states as function of the inter-cluster distance.

In the following, for the sake of the simplicity, we denote the set of the wave functions
obtained by the above-mentioned constrained energy variations as $\Phi_i^\pi$ where the subscript
$i$ is the index for each wave function. They are used as the basis wave functions for GCM
calculation explained below. 

\subsection{Angular momentum projection and\\ generator coordinate method}
After the energy variation, we project the wave function to the eigenstate of the angular
momentum. 
\begin{align}
 \Phi^{J^\pi}_{MKi} = n\frac{2J+1}{8\pi^2}\int d\Omega D^{J*}_{MK}(\Omega)R(\Omega)\Phi_i^\pi,
\end{align}
where $n$, $D^{J}_{MK}(\Omega)$ and $R(\Omega)$ are the normalization factor, Wigner's $D$
function and the rotation operator, respectively. Then, they are superposed to describe the
eigenstates of Hamiltonian,
\begin{align}
 \Psi^{J\pi}_{n} = \sum_{Ki} c_{Kin} \Phi^{J\pi}_{MKi}\label{eq:gcmwf}
 \end{align}
The coefficient $c_{Kin}$ is determined by solving Hill-Wheeler equation (GCM) 
\cite{Hill1953,Griffin1957},
\begin{align}
 &\sum_{K'j}\braket{\Phi^{J\pi}_{MKi}|H|\Phi^{J\pi}_{MK'j}}c_{K'jn}\nonumber\\
 &= E_n \sum_{K'j}\braket{\Phi^{J\pi}_{MKi}|\Phi^{J\pi}_{MK'j}}c_{K'jn}.
    \label{HillWheeler}
\end{align}
In the following, we call thus-obtained wave function GCM wave function. 

In general, a GCM wave function given by Eq. (\ref{eq:gcmwf}) is a mixture of various cluster
and non-cluster configurations. Therefore, we introduce two measures to identify cluster states
from the results of GCM calculations. The first measure is the overlap between the GCM wave
function and the basis wave function,  
\begin{align}
  O_i = |\langle \Psi^{J\pi}_n | \Phi^{J\pi}_i \rangle|^2.
  \label{def:GcmAmp}
\end{align}
If the overlap $O_i$ with a certain  $\Phi_i^{J\pi}$ is sufficiently large, the state described by
$\Psi_n^{J\pi}$ may be interpreted to have the cluster configuration described by
$\Phi_i^{J\pi}$. 

To define more quantitative measure, we introduce the projector to the cluster subspace. For
example, the projector to the subspace spanned by the $^{16}{\rm O}+{}^{12}{\rm C}$
configurations cluster is defined as
\begin{align}
  P_{^{16}{\rm O}+{}^{12}{\rm C}} &= \sum_{i}
 \ket{\Phi_{^{16}{\rm O}+{}^{12}{\rm C}}^{J^\pi}(d_j)}B^{-1}_{ij}
 \bra{\Phi_{^{16}{\rm O}+{}^{12}{\rm C}}^{J^\pi}(d_i)}, \\
 B_{ij} &= \braket{\Phi_{^{16}{\rm O}+{}^{12}{\rm C}}^{J^\pi}(d_i)|
 \Phi_{^{16}{\rm O}+{}^{12}{\rm C}}^{J^\pi}(d_j)}. 
  \label{def:SqOv}
\end{align}
Here, $\ket{\Phi_{^{16}{\rm O}+{}^{12}{\rm C}}^{J^\pi}(d)}$ denote the wave functions having
$^{16}{\rm O}+{}^{12}{\rm C}$ cluster configurations with inter-cluster distance $d$ obtained
by applying the $d$-constraint. The expectation value of $P_{^{16}{\rm O}+{}^{12}{\rm C}}$,
which we call cluster $S$-factor in the following, is a good measure to know to what extent a
GCM wave function is inside of  the $^{16}{\rm O}+{}^{12}{\rm C}$ cluster subspace;  
\begin{align}
 S_{^{16}{\rm O}+{}^{12}{\rm C}} = 
 \braket{\Psi_{n}^{J^\pi}|P_{^{16}{\rm O}+{}^{12}{\rm C}}|\Psi_{n}^{J^\pi}}.
\end{align}
The cluster $S$-factors for $^{24}{\rm Mg}+\alpha$, $^{20}{\rm Ne}+{}^{8}{\rm Be}$ and
$^{16}{\rm O}+{}^{12}{\rm C}$ configurations are also defined in the same manner. As already
explained above, when the inter-cluster distance $d$ is too small, the wave function obtained
by the $d$-constraint do not have cluster structure. Therefore, we use the wave functions
having non-small inter-cluster distance ($d\geq $ 4.0 fm) to define the projectors.

\section{Results and discussions}  \label{sec:3}
\subsection{Result of energy variation} \label{sec:3.1}
\begin{figure}
  \includegraphics[width=\hsize]{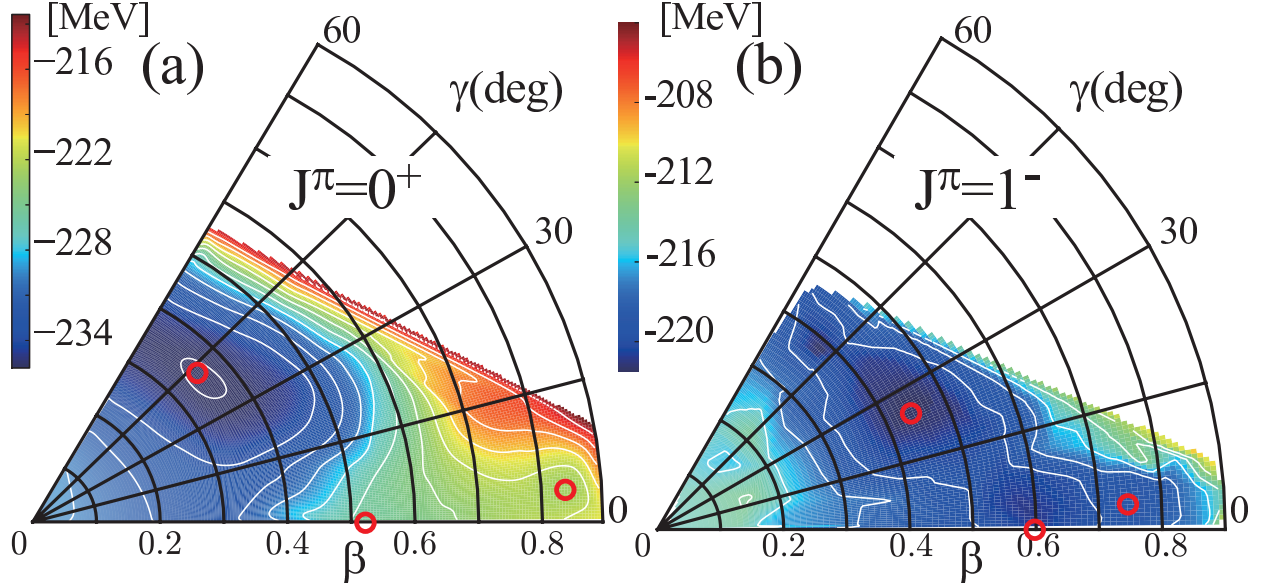}
  \caption{(Color online) The energy surfaces as functions of the quadrupole deformation
 parameters $\beta$ and $\gamma$ obtained by the energy variation with $\beta\gamma$-constraint
 and the angular momentum projection to the (a) $J^\pi=0^+$ and (b) $J^\pi=1^-$.
 Red circles show the energy minima or plateau.}
 \label{fig:surface_bg}
\end{figure}
\begin{figure}
  \includegraphics[width=1.0\hsize]{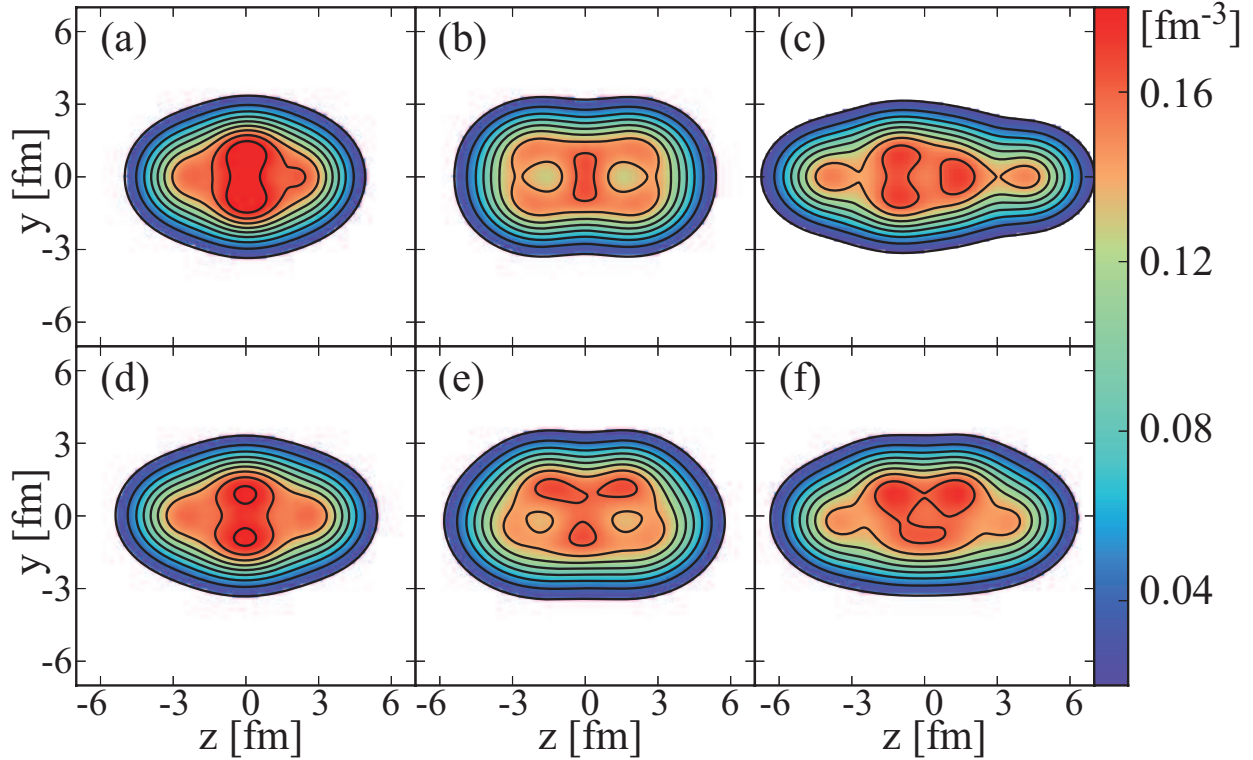}
 \caption{(Color online) Intrinsic matter density distributions of the minima on the  positive-
 and  negative-parity energy surfaces obtained by $\beta\gamma$-constraint shown in
 Fig. \ref{fig:surface_bg} .  The panels (a), (b) and (c) show the oblate, prolate and
 SD $J^\pi=0^+$ minima, while the panels (d),  (e) and (f) show the $J^\pi=1^-$ minima
 generated by the 1p1h  excitations from the  positive-parity minima. } 
 \label{fig:density_bg}
\end{figure}
The energy surface of the $J^\pi=0^+$ state obtained by the $\beta\gamma$ constraint and the
angular momentum projection is shown in Fig. \ref{fig:surface_bg} (a). There exist a couple of
energy minima or plateau with different quadrupole deformations. The lowest energy minimum has an
oblate shape of $(\beta,\gamma)=(0.36,46^\circ)$ with the energy $E=-235.7$ MeV, whose intrinsic
density distribution is shown in Fig. \ref{fig:density_bg} (a). Around this global minimum, the
energy surface is rather soft against both of  $\beta$ and $\gamma$ deformation. The second
lowest state is prolately  deformed as seen in its intrinsic density distribution
(Fig. \ref{fig:density_bg} (b)) and locates at $(\beta,\gamma)=(0.5,0^\circ)$ as a very shallow
energy minimum. Those two energy minima indicate the oblate and prolate shape coexistence in this
nucleus and yield the oblate deformed ground state and the prolate deformed $0^+_3$ state by
the GCM calculation, respectively. By further increase of the deformation, the third energy
minimum with strongly elongated shape (Fig. \ref{fig:density_bg} (c)) appears at
$(\beta,\gamma)=(0.85,5^\circ)$. As discussed in Ref. \cite{Taniguchi2009}, this configuration has
the a $(sd)^8(pf)^4$ configuration and becomes the dominant component of the $0^+_4$ state which
is regarded as the SD state.  

The energy surface of the negative-parity $J^\pi=1^-$ state shown in Fig. \ref{fig:surface_bg} (b) 
does not have clear local minima, but there are three shallow minima or plateau that are generated
by the single-particle excitations. The global minimum is located at 
$(\beta,\gamma)=(0.43,27^\circ)$ with the energy $E=-221.2$ MeV. This configuration has the
density distribution (Fig. \ref{fig:density_bg} (d)) similar to the oblate deformed ground state,
because it is generated by the one-nucleon excitation from the ground state configuration. 
As shown in Fig. \ref{fig:density_bg} (e) and (f), there also exist the prolate deformed and
SD negative-parity minima that are generated by the $1p1h$ excitations from the 
corresponding positive-parity minima. They are respectively located at 
$(\beta,\gamma)=(0.60,0^\circ)$ and $(0.73,2^\circ)$ with the energies $E=-220.8$ and -219.1
MeV. In terms of the Nilsson orbit, the prolate minimum is generated by the nucleon excitation
from the $[Nn_zm_l\Omega^\pi]=[211\ 1/2^+]$ orbit to the  $[330\ 1/2^-]$ orbit, while the
negative-parity SD minimum is generated by the nucleon deexcitation from the $[330\
1/2^-]$ orbit to the $[211\ 1/2^+]$ orbit.  
\begin{figure*}
\includegraphics[width=1.0\hsize]{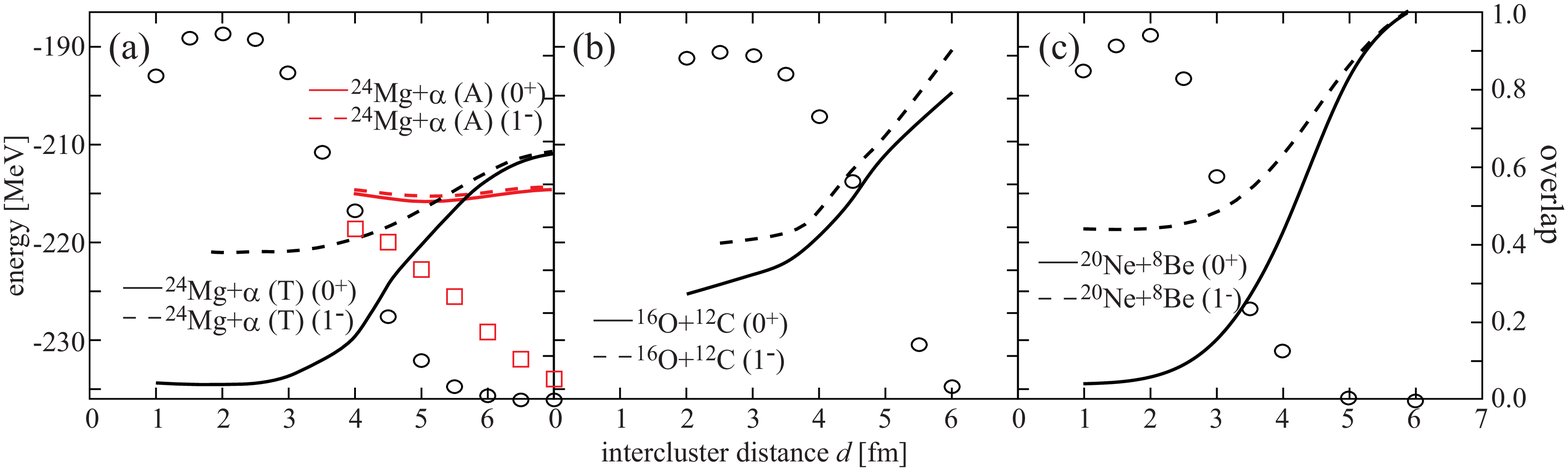}
\caption{Energy curves obtained by $d$-constraint. Panels (a), (b) and (c) respectively show
 the energy curves of the $J^\pi=0^+$ and $1^-$ states having ${}^{24}{\rm Mg}+\alpha$, 
 ${}^{16}{\rm  O}+{}^{12}{\rm C}$ and  ${}^{20}{\rm Ne}+{}^{8}{\rm Be}$ configurations. Circles
 and Boxes in the figure
 show the overlap between these $J^\pi=0^+$ cluster configurations and the energy minima on the
 $\beta\gamma$ energy surface. Black circles in panels (a) and (c) shows the overlap between the
 oblate deformed minimum (ground state) and $^{24}{\rm Mg}+\alpha {\rm (T)}$, 
 $^{20}{\rm  Ne}+^{8}{\rm Be}$  cluster configurations, while the red boxes in the panel (a)
 show the overlap between the SD minimum and $^{24}{\rm Mg}+\alpha {\rm (A)}$
 configuration. The circles in the panel (b) show the overlap between the prolate deformed
 minimum and the $^{12}{\rm C}+{}^{16}{\rm O}$  configuration.}
  \label{fig:surface_d}
\end{figure*}

\begin{figure}
  \includegraphics[width=1.0\hsize]{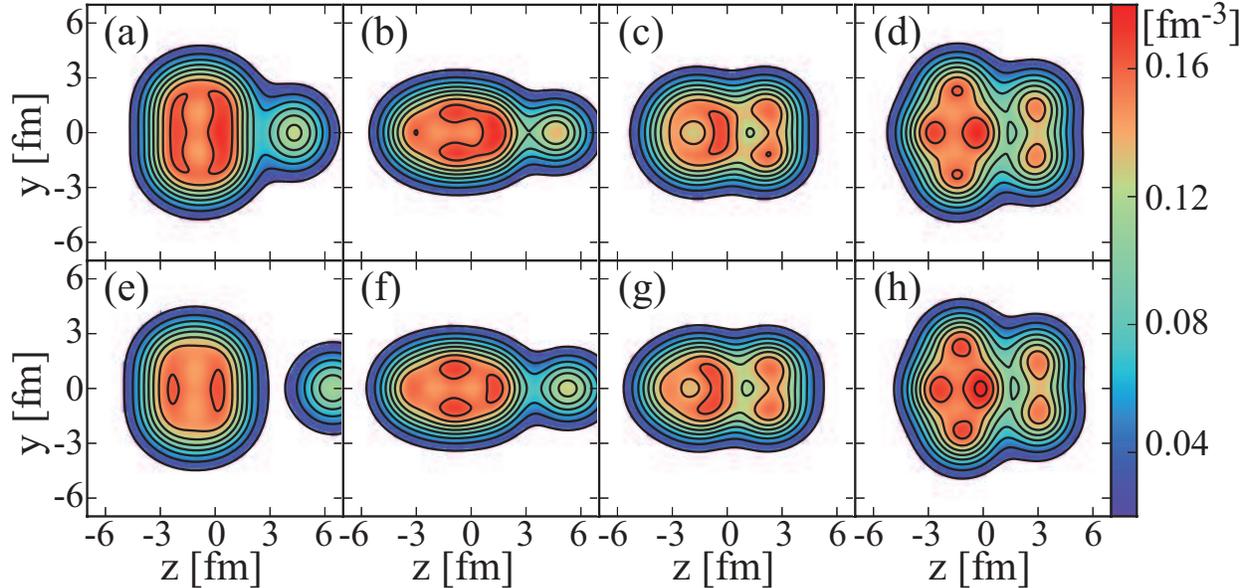}
 \caption{(Color online) Intrinsic matter density distributions obtained by $d$-constraint.
 (a), (b), (c) and (d) respectively show the ${}^{24}{\rm Mg}+\alpha$ (T), 
 ${}^{24}{\rm Mg}+\alpha$ (A), ${}^{16}{\rm O}+{}^{12}{\rm C}$ and 
 ${}^{20}{\rm Ne}+{}^{8}{\rm Be}$ configurations with $J^\pi=0^+$, while (e), (f), (g) and (h)
 show the $J^\pi=1^-$ partner having the same configurations.} 
 \label{fig:density_d}
\end{figure}

As confirmed from the intrinsic density distributions shown in Fig. \ref{fig:density_bg}, the
energy variation with the $\beta\gamma$ constraint does not generate prominent cluster
configurations, but mean-field configurations. On the other hand, the $d$-constraint yields
various cluster configurations. Figure \ref{fig:surface_d} shows the energy curves obtained
by the $d$-constraint. In the previous study, the $d$-constraint method was applied to the
positive-parity states of the $^{24}{\rm Mg}+\alpha$ and the $^{16}{\rm O}+{}^{12}$C
configurations. In addition to  them, in the present study, we applied it to the 
$^{20}{\rm Ne}+{}^{8}{\rm Be}$ configuration and investigated both of the positive- and
negative-parity. As a result, we found that all of these cluster configurations appear in both
parity states.

Figure \ref{fig:surface_d} (a) shows the energy curves for $^{24}{\rm Mg}+\alpha$ cluster
configurations projected to the $J^\pi=0^+$ and $1^-$ states. Because $^{24}{\rm Mg}$ cluster
is prolately deformed, two different $^{24}{\rm Mg}+\alpha$ configurations were obtained, in
which the orientations of $^{24}{\rm Mg}$ cluster are different. In the configuration denoted by
$^{24}{\rm Mg}+\alpha$ (T), the longest axis of $^{24}{\rm Mg}$ is perpendicular   to the
inter-cluster coordinate between $^{24}{\rm Mg}$ and $\alpha$ clusters. As a result, total system
is triaxially deformed as seen in its density distribution shown in Fig. \ref{fig:density_d}
(a). It is noted that, when the inter-cluster distance becomes small, the wave function of this
configuration with $J^\pi=0^+$ becomes almost identical to that of the oblate minimum on the
$\beta\gamma$ energy surface ({\it i.e.} the ground state). The overlap between the
wave functions of the $^{24}{\rm Mg}+\alpha$ (T) configuration and the oblate
deformed ground state minimum (circles in Fig. \ref{fig:surface_d} (a)) has the maximum value
0.96 at $d=2.0$ fm, and  their energies are very close to each 
other  (-235.7 MeV and -234.5 MeV,  respectively). Note that it does not necessarily mean that the
ground state is clustered, but it means the equivalence of the cluster and shell model wave
functions at small inter-cluster distance. This duality of shell and cluster is an essential
ingredient for the enhanced monopole and dipole transitions discussed in the section
\ref{sec:4}.

The negative-parity $J^\pi=1^-$ state with the same $^{24}{\rm Mg}+\alpha$ (T) configuration
(Fig. \ref{fig:density_d} (e))  appears at relatively high excitation energy that is approximately
15 MeV above the ground state. 
Different from the $J^\pi=0^+$ state, this negative-parity state has small overlap with the
negative-parity minima in the $\beta\gamma$ energy surface, which amount to 0.20 at most. Other
cluster configurations mentioned below also have  small overlap with the mean-field
configurations. This means that the $\beta\gamma$ constraint and $d$-constraint are describing the
different class of the negative-parity states. Namely, the $\beta\gamma$ constraint yields the
single-particle excited states built on the mean-filed, while the $d$-constraint yields the
reflection-asymmetric cluster states in which the relative motion between clusters have odd
angular momenta.

In another $^{24}{\rm Mg}+\alpha$ configuration denoted by $^{24}{\rm Mg}+\alpha$ (A), the
longest axis of the $^{24}{\rm Mg}$ cluster is parallel to the inter-cluster coordinate, and hence,
the system is axially deformed as shown in Fig. \ref{fig:density_d} (b). This configuration with
$J^\pi=0^+$ has large overlap with the SD configuration shown in
Fig. \ref{fig:density_bg} (c). The overlap between them amounts to 0.42 at the inter-cluster
distance $d=4.5$ fm. It is interesting to note that the negative-party $J^\pi=1^-$ state 
(Fig. \ref{fig:density_d} (f)) has almost the same intrinsic density distribution and almost the
same energy with the positive-parity $J^\pi=0^+$ state. It is because of the large inter-cluster
distance of the the $^{24}{\rm Mg}+\alpha$ (A) configuration compared to the
$^{24}{\rm Mg}+\alpha$ (T) configuration. The  $^{24}{\rm Mg}+\alpha$ (A) configuration in
negative-parity does not have corresponding state on $\beta\gamma$ energy surface having large
overlap.  

The $^{16}{\rm O}+{}^{12}{\rm C}$ configuration with $J^\pi=0^+$ appears approximately 10 MeV
above the ground state with the inter-cluster distance $d=2.5$ fm. At small inter-cluster
distance, this configuration has large overlap with the prolate minimum located at
$(\beta,\gamma)=(0.5,0^\circ)$ on the $\beta\gamma$ energy surface (Fig. \ref{fig:density_bg}
(c)). The overlap amounts to 0.90 at $d=2.5$ fm. The negative-parity $J^\pi=1^-$ state has similar
intrinsic density distribution to the positive-parity state as shown in Fig. \ref{fig:density_d}
(g) and has the excitation energy close to its positive-parity partner.  We also obtained another
$^{16}{\rm O}+{}^{12}{\rm C}$ configuration having different orientation of the 
$^{12}{\rm C}$ which may corresponds the highly excited $^{16}{\rm O}+{}^{12}{\rm C}$ cluster
states or molecular resonances 
\cite{Stokstad1972,Frohlich1976,Charles1976,James1976,Shawcross2001,Ashwood2001,Goasduff2014}.
However, its energy is rather high and is not discussed here.  Finally, we explain the
$^{20}{\rm Ne}+{}^{8}{\rm Be}$ configuration. It has the triaxial intrinsic density distribution
as shown in Fig. \ref{fig:density_d} (d) in which the longest axes of $^{20}{\rm Ne}$ and
$^{8}{\rm Be}$ clusters are parallel to each other, but perpendicular to the inter-cluster
coordinate. At small inter-cluster distance, this configuration with $J^\pi=0^+$ becomes identical
to the oblate minimum on $\beta\gamma$ energy surface (the ground state).  Thus, the ground state,
the $^{24}{\rm Mg}+\alpha$ and  $^{20}{\rm Ne}+{}^{8}{\rm Be}$ cluster configurations have large
overlap to each other at small inter-cluster distance. The negative-parity $J^\pi=1^-$ state of
$^{20}{\rm Ne}+{}^{8}{\rm Be}$ configuration appears at approximately 17 MeV above the ground
state with the inter-cluster distance $d=3.0$ fm.  

The result of the energy variation is summarized as follows. (1) The $\beta\gamma$ constraint
yielded three mean-field configurations with $J^\pi=0^+$ having oblate, prolate and SD
shapes. The oblate minimum has the lowest energy and corresponds to the ground state, while the
others constitute the excited $0^+$ states. (2) The $d$-constraint yielded prominent 
$^{24}{\rm Mg}+\alpha$ (T) and (A), $^{16}{\rm O}+{}^{12}{\rm C}$ and 
$^{20}{\rm Ne}+{}^{8}{\rm Be}$ cluster configurations with large inter-cluster distance and
smoothly connects them to the mean-filed states at small inter-cluster distance. These cluster
configurations have large overlap with the mean-field configurations indicating that the prolate,
oblate and SD minima on $\beta\gamma$ energy surface have duality of shell and 
cluster. Namely, the oblate deformed ground state has the duality of $^{24}{\rm Mg}+\alpha$ (T)
and $^{20}{\rm Ne}+{}^{8}{\rm Be}$ cluster configurations. The prolate minimum has the duality of
the $^{16}{\rm O}+{}^{12}{\rm C}$ configuration and the SD minimum has the duality of
the $^{24}{\rm Mg}+\alpha$ (A) configuration. (3) All of the cluster configurations are
accompanied by the negative-parity partner having almost the same intrinsic density
distributions. These negative-parity states originate in their reflection-asymmetric cluster
configurations.   

\subsection{Excitation spectrum and clustering} \label{3.2}
\begin{figure*}[t!]
  \includegraphics[width=\hsize]{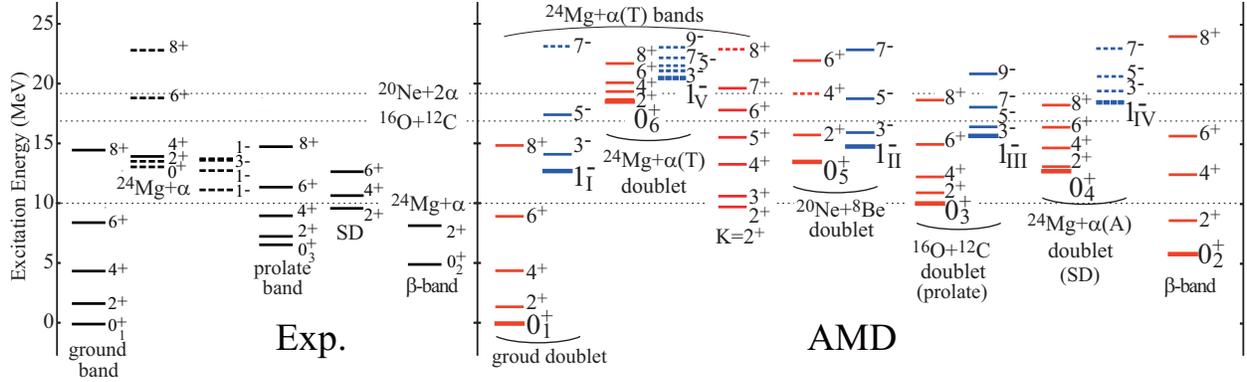}
  \caption{Calculated and observed partial level scheme of $^{28}{\rm Si}$. The levels shown by
 dashed lines are the averaged energies of the several states which have sizable cluster
 $S$-factors. Dotted lines show the $^{24}{\rm Mg}+\alpha$, $^{16}{\rm O}+{}^{12}{\rm C}$ 
 and $^{20}{\rm Ne}+2\alpha$ threshold energies.} 
  \label{fig:spectra}
\end{figure*}
\subsubsection{Overview of the spectrum}

Figure \ref{fig:spectra} shows the energy levels obtained by the GCM calculation together
with the corresponding observed states. In the figure, the rotational bands which have sizable
amount of the cluster $S$-factors and overlaps with the cluster wave functions are shown. Other
excited states are omitted except for the  $\beta$-band built on the ground-state band. The
detailed distribution of the cluster $S$-factors in the excited states is discussed in the section 
\ref{sec:4}. 

The results for the positive-parity states are consistent with our previous study, and we find
that the most of the positive-parity bands are accompanied by the negative-parity bands because of
their duality of mean-field and reflection-asymmetric clustering. 
The present result is briefly summarized as follows. The oblate minimum on the $\beta\gamma$
energy surface and the $^{24}{\rm Mg}+\alpha$ (T) configuration are mixed to each other and
generate a  group of the rotational bands denoted by $^{24}{\rm Mg}+\alpha$ (T) bands in
Fig. \ref{fig:spectra}. The oblate minimum also mixed with the $^{20}{\rm Ne}+{}^{8}{\rm Be}$
configuration to generate a pair of the positive- and negative-parity bands denoted by
$^{20}{\rm Ne}+{}^{8}{\rm Be}$ doublet. In a similar way, the prolate deformed minimum is mixed
with the $^{16}{\rm O}+{}^{12}{\rm C}$ configuration, and the SD minimum is mixed with
the $^{24}{\rm Mg}+\alpha$ (A) configuration. As a result, they respectively generate pairs of
the positive- and negative-parity bands, which are denoted by $^{16}{\rm O}+{}^{12}{\rm C}$ and
$^{24}{\rm Mg}+\alpha$ (A) doublets. In addition to them, the positive-parity band denoted by
$\beta$-band is generated by the $\beta$ vibration of the ground-state band.  

\subsubsection{$^{24}{Mg}+\alpha$ (T) bands}
A group of the rotational bands denoted by $^{24}{\rm Mg}+\alpha$ (T) bands includes three
positive-parity bands and two negative-parity bands. The positive-parity bands are the
ground-state band ($K^\pi=0^+$), $K^\pi=2^+$ band and another $K^\pi=0^+$ band built on the
$0^+_6$ state at 18.2 MeV. Because of the reflection-asymmetric clustering of the $^{24}{\rm
Mg}+\alpha$ (T) configuration, two negative-parity $K^\pi=0^-$ bands built on the $1^-_{\rm I}$
and $1^-_{\rm IV}$ are paired with the $K^\pi=0^+$ bands to constitute two parity doublets which 
are denoted by the ground doublet and  $^{24}{\rm Mg}+\alpha$ (T) doublet in the figure.

\begin{table}
 \caption{The calculated in-band $B(E2)$ ($e^2\rm fm^4$) strengths in the ground-state, 
 $^{24}{\rm Mg}+\alpha$ (T), $^{20}{\rm Ne}+{}^{8}{\rm Be}$, $^{16}{\rm O}+{}^{12}{\rm C}$,
 $^{24}{\rm Mg}+\alpha$ (A) doublets and in the  $\beta$-band. The numbers in the parenthesis
 are the experimental data \cite{Jenkins2012}.}  
 \label{tab:e2}
  \begin{ruledtabular}
  \begin{tabular}{cccc}
                           & ground & $^{24}{\rm Mg}+\alpha$ (T)&
                            $^{20}{\rm Ne}+{}^{8}{\rm Be}$ \\ \hline
  $B(E2;2^+\rightarrow0^+)$& 79.4 (67)   & 29.2   & 15.3    \\
  $B(E2;4^+\rightarrow2^+)$& 123 (83)    & 28.9   & 12.7    \\ \hline
  $B(E2;3^-\rightarrow1^-)$& 94.9        & 24.9   & 18.3    \\
  $B(E2;5^-\rightarrow3^-)$& 111         & 27.2   & 10.8    \\ \hline
   & $^{16}{\rm O}+{}^{12}{\rm C}$ & $^{24}{\rm Mg}+\alpha$ (A)& $\beta$-band \\ \hline
  $B(E2;2^+\rightarrow0^+)$& 221      & 664     & 52.8      \\
  $B(E2;4^+\rightarrow2^+)$& 299 (150)& 939     & 69.7      \\ \hline
  $B(E2;3^-\rightarrow1^-)$& 244      & 409     &           \\
  $B(E2;5^-\rightarrow3^-)$& 382      & 424     &           \\
  \end{tabular}
  \end{ruledtabular}
\end{table}

The ground-state band is dominated by the oblately deformed mean-field configuration shown in
Fig. \ref{fig:density_bg} (a) whose overlap with the GCM wave function amounts to 0.84. The 
moment-of-inertia of the ground-state band and the $B(E2)$ strengths listed in Tab. \ref{tab:e2} 
reasonably agree with the observed data, indicating that the deformed mean-field nature of the
ground state is properly described by the present calculation. However, it must be noted that
this ground-state band also has the large overlap with the $^{24}{\rm Mg}+\alpha$ (T)
configuration with small inter-cluster distance, which amount to 0.96 for $d=2.0$ fm. This means
that the 
ground-state band has a duality of the oblate shaped mean-field and $^{24}{\rm Mg}+\alpha$ (T)
clustering. Therefore, the excitation of the inter-cluster motion between $^{24}{\rm Mg}$ and
$\alpha$ clusters yields excited bands with prominent clustering. The $1\hbar\omega$ excitation
of the inter-cluster motion yields the $K^\pi=0^-$ band built on the $1^-$ state denoted by
$1^-_{\rm I}$ which is dominated by the negative-parity $^{24}{\rm Mg}+\alpha$ (T) configuration
with $d=2.5$ fm. Hence, we assigned it as the partner of the ground-state band which constitutes the
ground doublet, although the $1p1h$ mean-field configuration (Fig. \ref{fig:density_bg} (d)) also
has non-negligible contribution to this band. In  addition to the ground doublet, the 2 and
$3\hbar\omega$ excitations of the inter-cluster motion yield $K^\pi=0^\pm$ bands built on the
$0^+_6$ state and a group of $1^-$ states denoted by $1^-_{\rm V}$, which constitute another
doublet  denoted by $^{24}{\rm Mg}+\alpha$ (T)  doublet. Because of the coupling with other
non-cluster configurations, three $1^-$ states around 22 MeV have large $^{24}{\rm Mg}+\alpha$
cluster $S$-factors, and their averaged energy is denoted by $1^-_{\rm V}$ in
Fig. \ref{fig:spectra}.  As the inter-cluster motion is largely excited, these bands have
prominent cluster structure. The $0^+_6$ and $1^-_{\rm V}$ states have large overlap with the
configurations shown in Fig. \ref{fig:density_d} (a) and (e) which amount to 0.32 and 0.22
(averaged), respectively.  In addition to these parity doublets,  the triaxial deformation of the
$^{24}{\rm Mg}+\alpha$ (T) configuration yields $K^\pi=2^+$ band. Thus, the duality of the ground
state yields two parity doublets and $K^\pi=2^+$ band which  are classified as 
$^{24}{\rm Mg}+\alpha$ (T) bands.

\subsubsection{$^{20}{Ne}+{}^{8}{\rm Be}$ doublet and $\beta$-band}
As discussed in the section \ref{sec:3.1}, the oblate deformed minimum also has large overlap
with the $^{20}{\rm Ne}+{}^{8}{\rm Be}$ configuration with small inter-cluster distance. Therefore,
the excitation of the inter-cluster motion between $^{20}{\rm Ne}$ and $^{8}{\rm Be}$ clusters
should yield a series of the cluster bands. In addition, the deformed mean-field aspect
of the ground state can yield a different kind of excitation mode, {\it i.e.} the $\beta$ vibration.
The GCM calculation showed that these two excitation modes strongly mix to each other to yield the 
$0^+_2$ and $0^+_5$ states. Their overlap with the configurations shown in Fig. \ref{fig:density_d}
(d) amount to 0.13 and 0.18, respectively. Similarly to the $^{24}{\rm Mg}+\alpha$
configuration,  the $^{20}{\rm Ne}+{}^{8}{\rm Be}$ configuration yields the $1^-_{\rm II}$ state
which is paired with the $0^+_5$ state to constitute the parity doublet denoted by
$^{20}{\rm Ne}+{}^{8}{\rm Be}$ doublet.

\subsubsection{$^{16}{ O}+{}^{12}{ C}$ doublet and $^{24}{ Mg}+\alpha$ (A) doublet}
In addition to the above-mentioned bands related to the ground state duality, there are other
bands which are unrelated to the ground state. The prolate band is built on the $0^+_3$ state at
10.0 MeV which has the large overlap with the prolate deformed local minimum shown in
Fig. \ref{fig:density_bg} (b). Combined with the oblately deformed ground 
state, this prolate deformed $0^+_3$ state indicates the shape coexistence in the low-lying states
of  $^{28}{\rm Si}$. As discussed in the previous work \cite{Taniguchi2009}, this prolate band has
large overlap with the $^{16}{\rm O}+{}^{12}{\rm C}$ cluster configuration. Hence, it is concluded
that the prolate band has the duality of the prolate deformed mean-field and 
$^{16}{\rm O}+{}^{12}{\rm C}$ clustering. The negative-parity band built on the $1^-_{\rm III}$
state also has the large overlap with the $^{16}{\rm O}+{}^{12}{\rm C}$ configuration 
(Fig. \ref{fig:density_d} (g)), and assigned as the partner of the positive-parity prolate band, 
that constitutes the $^{16}{\rm O}+{}^{12}{\rm C}$ doublet.

Another prolate deformed minimum {\it i.e.} the SD minimum located at
$(\beta,\gamma)=(0.85,5^\circ)$ generates the SD band built on the $0^+_{4}$ state at 12.7
MeV. This band has large overlap with the $^{24}{\rm Mg}+\alpha$ (A) configuration shown in 
the Fig. \ref{fig:density_d} (f). There are two $1^-$ states having large overlap with the
negative-parity $^{24}{\rm Mg}+\alpha$ (A) configuration and the mixing with the 
$^{16}{\rm O}+{}^{12}{\rm C}$ configuration.  Their averaged energy is denoted by $1^-_{\rm IV}$
and the negative-parity band built on these states is associated with the positive-parity band. 
We denote this doublet as $^{24}{\rm Mg}+\alpha$ (A) doublet. As discussed in the previous
section, the $^{16}{\rm O}+{}^{12}{\rm C}$ and  ${}^{24}{\rm Mg}+\alpha$ (A) configurations do not
have overlap with the oblate deformed ground state. Therefore, the $^{16}{\rm O}+{}^{12}{\rm C}$
and  ${}^{24}{\rm Mg}+\alpha$ (A) doublets are disconnected with the ground state.

\subsubsection{Systematics of clustering and observed candidates}

\begin{figure*}[t!]
  \includegraphics[width=0.7\hsize]{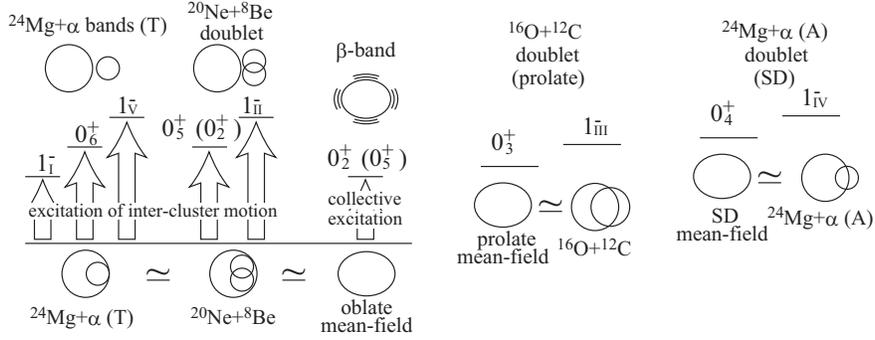}
  \caption{The oblate deformed ground state has the duality of the mean field and 
 $^{24}{\rm  Mg}+\alpha$ (T) and $^{16}{\rm O}+{}^{12}{\rm C}$ clustering. From the duality,
 $\beta$-band, $^{24}{\rm Mg}+\alpha$ bands and $^{20}{\rm Ne}+{}^{8}{\rm Be}$ doublet arise.}
  \label{fig:illust1}
\end{figure*}

To summarize this section, the systematics of the cluster states explained above is
schematically illustrated in Fig. \ref{fig:illust1}. The ground state has duality of the oblate
deformed mean-field,  $^{24}{\rm Mg}+\alpha$ (T) and $^{20}{\rm Ne}+{}^{8}{\rm Be}$ cluster
configurations. By the excitation of the inter-cluster motion between $^{24}{\rm Mg}$ and $\alpha$
clusters yields a group of  the $^{24}{\rm Mg}+\alpha$ (T) bands. The $^{20}{\rm Ne}+{}^{8}{\rm Be}$
clustering also arise from the ground state duality and it is strongly mixed with the
$\beta$-vibration mode which arises from the mean-field aspect of the ground state. Aside from
these bands, the $^{16}{\rm O}+{}^{12}{\rm C}$ and  ${}^{24}{\rm Mg}+\alpha$ (A) doublets exist
and disconnected from the ground state, because of the orthogonality of their cluster
configurations to the ground state.  

Experimentally, the low-lying three positive-parity bands, {\it i.e.} the ground-state band,
$\beta$-band and the prolate band, are assigned firmly and coincide with the present
calculation, although the calculation slightly overestimates the energies of the excited bands.
On the other hand, the experimental assignment of the negative-parity bands and high-lying bands
are not established yet, and hence, the assignment of the cluster bands is still ambiguous. 
Many experiments have been performed to identify the cluster bands
\cite{Jenkins2012,Maas1978,Cseh1982,Tanabe1983,Kubono1986,Artemov1990,Stokstad1972,Goasduff2014},
and Fig. \ref{fig:spectra} shows the candidates of the cluster bands reported in Refs. 
\cite{Jenkins2012,Cseh1982,Artemov1990}, which energetically coincide with the present
calculation. A couple of $0^+, 2^+$ and $1^-, 3^-$ states were reported around $E_x=13$ MeV by the
$\alpha$ transfer and radiative $\alpha$ capture reactions. They have relatively large
$\alpha$ decay width, hence, can be regarded as the candidates of the $^{24}$Mg+$\alpha$ (T) or
$^{24}$Mg+$\alpha$ (A) doublets. Furthermore, based on the analysis of the 
$^{24}{\rm Mg}(\alpha,\gamma)$ and  $^{12}{\rm C}(^{20}{\rm Ne},\alpha)^{28}{\rm Si}$ reactions,
another rotational band was suggested \cite{Jenkins2012} and the authors were assigned it to the
SD band predicted by the previous AMD study \cite{Taniguchi2009}.

\section{isoscalar monopole and dipole transitions} \label{sec:4}
Here, we discuss that part of the clustering systematics summarized above can be detected by the  
IS monopole and dipole transitions from the ground state. To illustrate it, we first
discuss the the duality of shell and cluster. Then, we present the result of AMD calculation
to show that the IS monopole and dipole transitions strongly yield cluster states.

\subsection{Duality of shell and cluster} \label{sec:4.1}
In the section \ref{sec:3}, we have explained that the many low-lying positive-parity states
have the duality of the mean-field (shell) and cluster. Here, we show that it is reasonably
understood by the $SU(3)$ shell model \cite{Elliott1958,Elliott1958a} and the Bayman-Bohr theorem 
\cite{Perring1956,Bayman1958}.  

$^{28}{\rm Si}$ has 12 nucleons in $sd$-shell on top of the $^{16}{\rm O}$ core and its oblate
deformed ground state can be approximated by the $(\lambda,\mu)=(0,12)$ representation of $SU(3)$
shell  model. Denoting the eigenstates of three-dimensional harmonic oscillator in the Cartesian
representation $(n_xn_yn_z)$, it is written as
\begin{align}
 (\lambda,\mu)=(0,12) :\ (002)^4(011)^4(020)^4,
\end{align}
where the configuration of 12 nucleons are explicitly shown, and the $^{16}{\rm O}$ core 
which corresponds to $(000)^4(100)^4(010)^4(001)^4$ is omitted. In a same manner, the prolate
deformed state is approximated by the $(\lambda,\mu)=(12,0)$ representation,
\begin{align}
 (\lambda,\mu)=(12,0) :\ (002)^4(011)^4(101)^4, \label{eq:012}
\end{align}
in which the orbit occupied by the last four nucleons is different from that in the
ground state. The excitation of the last four nucleons into $pf$-shell yields the SD configuration
which is given by the $(\lambda,\mu)=(20,4)$ representation.
\begin{align}
 (\lambda,\mu)=(20,4) :\ (002)^4(011)^4(003)^4. \label{eq:su204}
\end{align}

\begin{figure}[t!]
  \includegraphics[width=0.8\hsize]{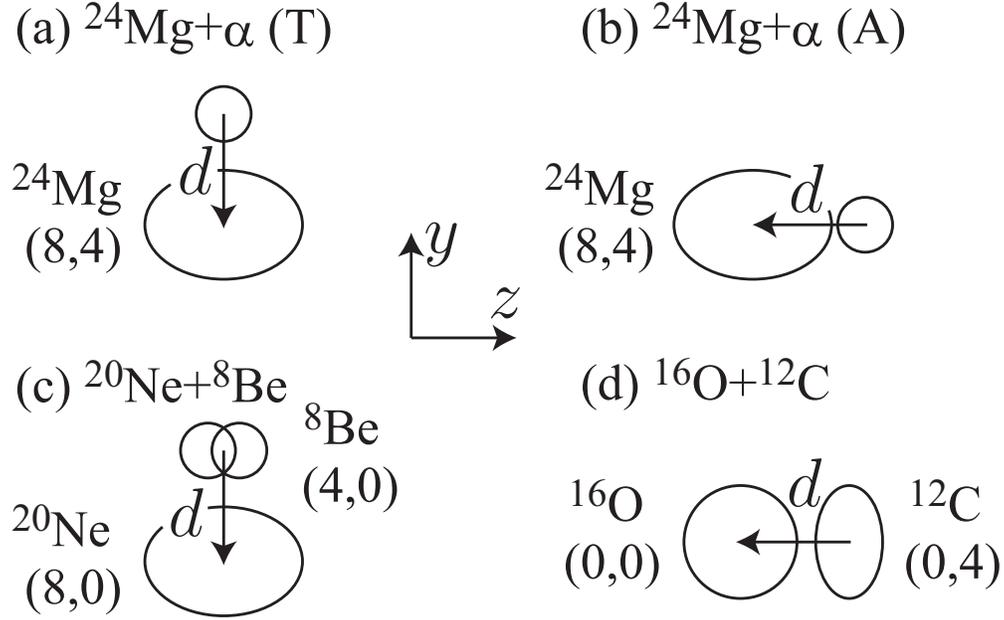}
  \caption{Schematic figure for the various cluster configurations and their duality.}
  \label{fig:illust2}
\end{figure}

The Bayman-Bohr theorem tells that these $SU(3)$ shell model wave functions are identical to the
cluster model wave functions with the zero inter-cluster distance. First, we consider 
${}^{24}{\rm Mg}+\alpha$ cluster configurations. The triaxially deformed ground state of
${}^{24}{\rm Mg}$ is given by the $(\lambda,\mu)=(8,4)$ representation,
\begin{align}
 (\lambda,\mu)=(8,4) :\ (002)^4(011)^4,
\end{align}
where the longest deformation axis is $z$-axis. Now we consider the ${}^{24}{\rm Mg}+\alpha$
configuration in which the $\alpha$ cluster is placed on the $y$-axis as illustrated in
Fig. \ref{fig:illust2} (a). This configuration corresponds to the $^{24}{\rm Mg}+\alpha$ (T)
configuration obtained by AMD calculation. Since the orbits $(000)$ and $(010)$ are already
occupied by the nucleons in $^{24}{\rm Mg}$ cluster, the nucleons in the $\alpha$ cluster having
$(000)^4$ configuration must occupy $(020)^4$ at zero inter-cluster distance $d=0$. As a result,
one sees that the $^{24}{\rm Mg}+\alpha$ (T) cluster configuration becomes identical to the
$(0,12)$ representation for the ground state given by Eq. (\ref{eq:012}), 
\begin{align}
 \lim_{d\rightarrow 0} \Phi_{^{24}{\rm Mg}+\alpha ({\rm T})}(d) = (002)^4(011)^4(020)^4.
\end{align}
This clearly explains why the $^{24}{\rm Mg}+\alpha$ (T) configuration obtained by the
$d$-constraint becomes almost identical to the ground state configuration at small inter-cluster
distance $d$. The placement of the $\alpha$ particle on $z$-axis corresponds the another
configuration $^{24}{\rm Mg}+\alpha$ (A) as shown in Fig. \ref{fig:illust2} (b). In this case, at
zero inter-cluster distance, the nucleons in the $\alpha$ cluster occupy $(003)^4$ resulting in
the SD configuration given by Eq. (\ref{eq:su204}), 
\begin{align}
 \lim_{d\rightarrow 0} \Phi_{^{24}{\rm Mg}+\alpha ({\rm A})}(d) = (002)^4(011)^4(003)^4.
\end{align}

The other cluster configurations are also considered in the same way. The prolately deformed
ground state of $^{20}{\rm Ne}$ is given by the $(\lambda,\mu)=(8,0)$ representation,
\begin{align}
 (\lambda,\mu)=(8,0) :\ (002)^4,
\end{align}
where the symmetry axis of $^{20}{\rm Ne}$ is $z$-axis. The $^{20}{\rm Ne}+{}^{8}{\rm Be}$
configuration corresponds to the placement of $^{8}{\rm Be}$ cluster on $y$-axis where the
symmetry axis of $^{8}{\rm Be}$ is also $z$-axis (Fig. \ref{fig:illust2} (c)).  At zero
inter-cluster distance, the nucleons in the $^{8}{\rm Be}$ cluster having $(000)^4(001)^4$
configuration occupy $(020)^4(011)^4$ due to the Pauli principle, and one finds it is identical
to the ground state configuration,
\begin{align}
 \lim_{d\rightarrow 0} \Phi_{^{20}{\rm Ne}+{}^{8}{\rm Be} }(d) = (002)^4(011)^4(020)^4.
\end{align}
The oblately deformed ground state of $^{12}{\rm C}$ is given by the $(\lambda,\mu)=(0,4)$
representation, 
\begin{align}
 (\lambda,\mu)=(0,4) :\ (000)^4(100)^4(010)^4,
\end{align}
where the symmetry axis is $z$-axis. The $^{16}{\rm O}+{}^{12}{\rm C}$ configuration corresponds
the placement of $^{16}{\rm O}$ and ${}^{12}{\rm C}$ clusters on $z$-axis as shown in
Fig. \ref{fig:illust2} (d). At zero inter-cluster distance, it is identical to the prolate
deformed state,
\begin{align}
 \lim_{d\rightarrow 0} \Phi_{^{16}{\rm O}+{}^{12}{\rm C} }(d) = (002)^4(011)^4(101)^4.
\end{align}
Thus, considering the corresponding $SU(3)$ shell model wave function, the duality of the 
mean-field and cluster configurations illustrated in Fig. \ref{fig:illust1} is clearly explained.
It is also noted that the duality of the shell and cluster in $^{28}{\rm Si}$  was also
investigated and found by the Skyrme Hartree-Fock calculation \cite{Maruhn2006}. 

As discussed in Refs. \cite{Yamada2008,Horiuchi2012,Chiba2016}, this duality of
shell and  cluster means that the degree-of-freedom of ${}^{28}{\rm Mg}+\alpha$ (T) and
$^{20}{\rm Ne}+{}^{8}{\rm Be}$ cluster excitations are embedded in the ground state. Therefore,
the excitation of the inter-cluster motion embedded in the ground state yields excited cluster
states with pronounced ${}^{24}{\rm Mg}+\alpha$ (T) and ${}^{20}{\rm Ne}+{}^{8}{\rm Be}$ 
configurations. The important fact is that the IS monopole and dipole transitions between the
ground state and these excited cluster states are very strong, and hence, these transitions are
very good probe for the clustering. Indeed, if one assumes that the ground state is a pure
$SU(3)$ shell model state and the excited cluster states are described by the cluster model wave
function, it is possible to analytically show the enhancement of the IS monopole and dipole
transitions \cite{Yamada2008,Chiba2016}. However, in the case of $^{28}{\rm Si}$, the ground state
deviates from a pure $SU(3)$ shell model state because of the strong influence of the spin-orbit
interaction.  In addition to this, the coupling between the cluster configurations and mean-field 
configurations in the excited states is not negligible. Therefore, the evaluation of the
transition strengths by the realistic nuclear model is indispensable for the quantitative
discussions. For this purpose, we present the results of the GCM calculation below.

\subsection{Isoscalar monopole and dipole transitions}\label{sec:4.2}
Using the wave functions of the ground and excited cluster states obtained by the GCM
calculation, the IS monopole and dipole transition strengths are directly evaluated.
The transition operators and matrix elements between the ground and excited states are given
as,
\begin{align}
 {\mathcal M}(IS0) &= \sum_{i=1}^A(\bm r_i - \bm r_{cm})^2,\\
 {\mathcal M}_\mu(IS1) &= \sum_{i=1}^A(\bm r_i - \bm r_{cm})^2
 {\mathcal Y}_{1\mu}(\bm r_i - \bm r_{cm})\\
 M(IS0;0^+_1\rightarrow 0^+_n) &= \braket{0^+_n|{\mathcal M}(IS0)|0^+_1},\\
 M(IS1;0^+_1\rightarrow 1^-_n) &= \braket{1^-_n||{\mathcal M}_0(IS1)||0^+_1}\nonumber\\
 &=\sqrt{3}\braket{1^-_n|{\mathcal M}_0(IS1)|0^+_1},
\end{align}
where $\bm r_i$ and $\bm r_{cm}$ denote the single-particle and center-of-mass coordinates, 
respectively. The solid spherical harmonics is defined as
${\mathcal Y}_{1\mu}(\bm r) = rY_{1\mu}(\hat r)$.
We also calculated the cluster $S$-factors of the ground and exited states to
see how the clustering and IS monopole and dipole transitions are correlated to each other.

\begin{figure*}[t!]
 \includegraphics[width=\hsize]{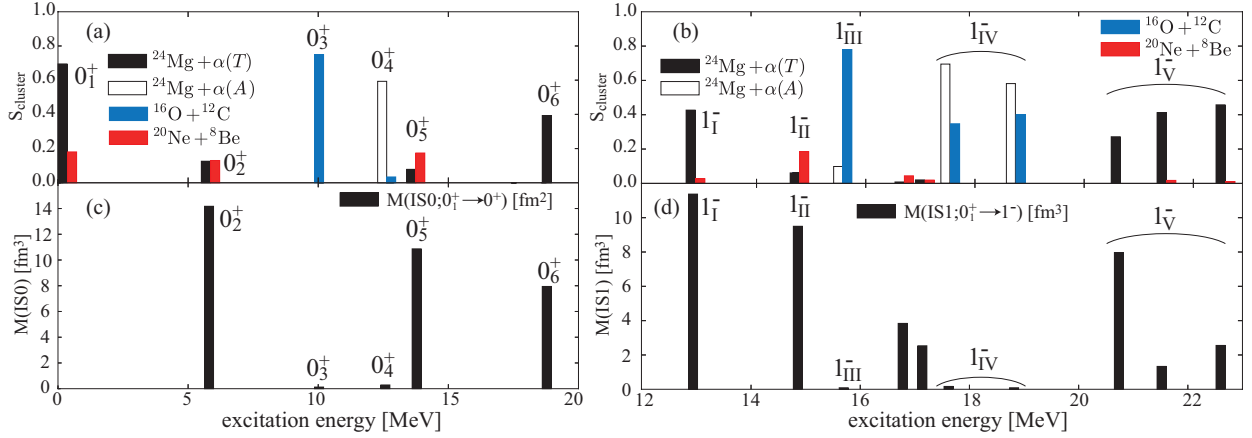}
 \caption{(a) The cluster $S$-factors for the ground and excited $0^+$ states. (b) Same as (a) but
 for the excited $1^-$ states. (c) The IS monopole transition matrix from the ground state to the
 excited $0^+$ states. (d) The IS dipole transition matrix from the ground state to the
 excited $1^-$ states.}
 \label{fig:is01}
\end{figure*} 
 
The results are shown in Fig. \ref{fig:is01} where the panels (a) and (b) show the cluster
$S$-factors of $0^+$ and $1^-$ states, while the panels (c) and (d) shows the IS monopole
and dipole transition matrices. The calculated $S$-factors confirm the clustering
systematics summarized in Fig. \ref{fig:illust1}. Owing to the duality of the shell and
cluster, the ground state has large $S$-factor for the $^{24}{\rm Mg}+\alpha$ (T)
configuration which amounts to 0.68 and has non-negligible $S$-factor for the 
$^{20}{\rm Ne}+{}^{8}{\rm Be}$ configuration. The $0^+_6$ state is regarded as the
pronounced $^{24}{\rm Mg}+\alpha$ (T) cluster state, while the $0^+_5(0^+_2)$ state
is regarded as the pronounced $^{20}{\rm Ne}+{}^{8}{\rm Be}$ cluster states from their
$S$-factors. One also sees that the $1^-_{\rm I}$ state and a group of $1^-$ state denoted
by $1^-_{\rm V}$ are the  $^{24}{\rm Mg}+\alpha$ (T) cluster states and paired with the ground
state and the $0^+_6$ state, respectively, while the $1^-_{\rm II}$ state should be paired
with the $0^+_5$ state to constitute the $^{20}{\rm Ne}+{}^{8}{\rm Be}$ doublet.
In the same way, the $S$-factors clearly show that the $0^+_3$ and $1^-_{\rm III}$ states
are the $^{16}{\rm O}+{}^{12}{\rm C}$ doublet, and the $0^+_4$ and $1^-_{\rm IV}$ states are
the $^{24}{\rm Mg}+\alpha$ (A) doublet although the ${}^{16}{\rm O}+{}^{12}{\rm C}$ 
configuration is also mixed in the $1^-_{\rm IV}$ states.

It is very impressive to see that the IS monopole and dipole strengths shown in the 
panels (c) and (d) are very strongly correlated to the $S$-factors for the 
${}^{24}{\rm Mg}+\alpha$ (A) and $^{20}{\rm Ne}+{}^{8}{\rm Be}$ configurations, but almost 
insensitive to the other cluster and non-cluster states except for a couple of $1^-$ states
around 17 MeV. Because the IS monopole and dipole operators activate the degree-of-freedom
of cluster excitation embedded in the ground state, the transition strengths to the 
${}^{24}{\rm Mg}+\alpha$ (A) and $^{20}{\rm Ne}+{}^{8}{\rm Be}$ cluster states are as strong 
as the single-particle estimates which are given as,
\begin{align}
 M(IS0)_{WU} &= \frac{3}{5}(1.2A^{1/3})^2 \simeq 8.0\ {\rm fm^2}\\ 
 M(IS1)_{WU} &= \sqrt{\frac{3}{16\pi}}(1.2A^{1/3})^3 \simeq 11.8\ {\rm fm^2}.
\end{align}
Therefore, we can conclude that the IS monopole and dipole transitions are good probe to identify 
the $^{24}{\rm Mg}+\alpha$ (T) and $^{20}{\rm Ne}+{}^{8}{\rm Be}$ clustering.

Recently, an interesting and promising experimental data was reported by the measurement of the
$^{28}{\rm Si}(\alpha,\alpha'){}^{28}{\rm Si}^*$ inelastic scattering 
\cite{Peach2016,Adsley2016}. It was found that a couple of $0^+$ states above 9 MeV are strongly
populated and deduced to have large IS monopole transition strengths. Hence, they are suggested as
strong candidates of the $\alpha$  cluster states \cite{Adsley2016}. We expect that the detailed
comparison of the IS monopole and dipole transition strengths between experiment and theory will
reveal the clustering systematics in $^{28}{\rm Si}$. 

\section{Summary}  \label{sec:5}
In summary, we have investigated the clustering systematics in $^{28}{\rm Si}$ based on the 
antisymmetrized molecular dynamics. It is found that the inversion doublet bands with various
kinds of reflection-asymmetric cluster configuration appears in the excited states, and the IS
monopole and dipole transitions are good probe for ${}^{24}{\rm Mg}+\alpha$ (T) and
$^{20}{\rm Ne}+{}^{8}{\rm Be}$ cluster states.

The energy variation by using $d$-constraint yielded various kinds of cluster configurations with
positive- and negative-parity, while the $\beta\gamma$-constraint yielded mean-field
configurations. It is found that the cluster configurations become identical to the mean-field
configurations at small inter-cluster distance because of the duality of mean-field (shell) and
cluster. In particular, it is emphasized that the oblate deformed  ground state has the duality of  
the ${}^{24}{\rm Mg}+\alpha$ (T) and $^{20}{\rm Ne}+{}^{8}{\rm Be}$ configurations. 

The GCM calculation showed that a group of the ${}^{24}{\rm Mg}+\alpha$ (T) and
$^{20}{\rm Ne}+{}^{8}{\rm Be}$ cluster bands are generated by the excitation of the 
inter-cluster motion embedded in the ground state. In addition to them, the prolate and SD
bands have the duality of $^{16}{\rm O}+{}^{12}{\rm C}$ and $^{20}{\rm Ne}+{}^{8}{\rm Be}$
clustering, respectively. Because of their reflection-asymmetric intrinsic configurations, they
are accompanied by the negative-parity bands to constitute the inversion doublets.

Because of the duality of the ground state, it is numerically shown that the 
${}^{24}{\rm Mg}+\alpha$ (T) and $^{20}{\rm Ne}+{}^{8}{\rm Be}$ cluster bands have enhanced
IS monopole and dipole transition matrices which are as large as the single-particle estimates. 
On the other hand, other cluster states and non-cluster states are rather insensitive to the
IS monopole and dipole transitions. Hence, we conclude that the ${}^{24}{\rm Mg}+\alpha$ (T)
and $^{20}{\rm Ne}+{}^{8}{\rm Be}$ cluster bands can be identified from their enhanced 
transitions. We expect that more quantitative comparison with the experiments will reveal
the clustering systematics in $^{28}{\rm Si}$.

\begin{acknowledgements}
The authors thanks to Prof. Kanada-En'yo and Prof. Kawabata and Prof. Ito for the fruitful
discussions. Part of the numerical calculations were performed by using the supercomputers at 
the High Energy Accelerator Research Organization (KEK) and at Yukawa Institute for Theoretical
 Physics (YITP) in Kyoto University. The support by the Grants-in-Aid for Scientific Research on
 Innovative Areas from MEXT (Grant No. 2404:24105008) and JSPS KAKENHI Grant Nos. 16J03654,
 25800124 and  16K05339 are acknowledged.
\end{acknowledgements}

\bibliography{Si28paper}

\end{document}